\documentclass[conference]{IEEEtran}

\usepackage{cite}
\usepackage{amsmath,amssymb,amsfonts}
\usepackage{graphicx}
\usepackage{textcomp}
\usepackage{xcolor}
\usepackage{changes}
\usepackage{amsthm}
\usepackage{algorithm}      
\usepackage{algpseudocode}  
\usepackage{booktabs}
\usepackage{multirow}
\graphicspath{ {./images/} }

\usepackage{array}   
\usepackage{booktabs} 
\usepackage{makecell}
\usepackage{tabularx}

\title{Unsupervised Dataset Cleaning Framework for Encrypted Traffic Classification }

\author{
\IEEEauthorblockN{Kun Qiu\IEEEauthorrefmark{1}, Ying Wang\IEEEauthorrefmark{2}, Baoqian Li\IEEEauthorrefmark{2}, Wenjun Zhu\IEEEauthorrefmark{1}}
\IEEEauthorblockA{\IEEEauthorrefmark{1}Fudan University, Shanghai, China
\IEEEauthorblockA{\IEEEauthorrefmark{2}Intel Asia-Pacific Research \& Development Ltd, Shanghai, China\\
\{qkun, wenjun\}@fudan.edu.cn}\{ying.a.wang, baoqian.li\}@intel.com}
}

\begin{document}
\maketitle

\begin{abstract}
Traffic classification, a technique for assigning network flows to predefined categories, has been widely deployed in enterprise and carrier networks. With the massive adoption of mobile devices, encryption is increasingly used in mobile applications to address privacy concerns. Consequently, traditional methods such as Deep Packet Inspection (DPI) fail to distinguish encrypted traffic. To tackle this challenge, Artificial Intelligence (AI)—in particular Machine Learning (ML)—has emerged as a promising solution for encrypted traffic classification. A crucial prerequisite for any ML-based approach is traffic data cleaning, which removes flows that are not useful for training (e.g., irrelevant protocols, background activity, control-plane messages, and long-lived sessions). Existing cleaning solutions depend on manual inspection of every captured packet, making the process both costly and time-consuming. In this poster, we present an unsupervised framework that automatically cleans encrypted mobile traffic. Evaluation on real-world datasets shows that our framework incurs only a $2\% \sim 2.5\%$ reduction in classification accuracy compared with manual cleaning. These results demonstrate that our method offers an efficient and effective preprocessing step for ML-based encrypted traffic classification.
\end{abstract}
\begin{IEEEkeywords}
Traffic Analytics, Traffic Cleaning, Unsupervised Learning
\end{IEEEkeywords}

\section{Introduction}
\label{sec:intro}
Network traffic classification assigns flows to categories such as specific applications or benign and malicious behavior~\cite{pacheco2018towards,marnerides2014traffic,salau2024software}. Traditional techniques rely on IANA port numbers or deep packet inspection to examine payload contents~\cite{xu2023harry,xu2024accelerating}. As mobile applications increasingly deploy encryption like TLS~\cite{aceto2019mobile}, these methods can no longer distinguish among different application flows.

Machine learning has emerged as a powerful solution for encrypted traffic classification~\cite{azab2024network,sharma2024review}, but it depends on clean training data. Incomplete handshakes, background service flows, control plane messages and long‐lived sessions must be removed before model training~\cite{bakhshi2016internet,usama2019unsupervised}. Gathering mobile traffic at routers or edge devices inevitably brings in many unwanted flows, and manually filtering every packet can take days, creating a major barrier to practical deployment~\cite{divakaran2015slic}.

We present an online framework that automates encrypted traffic cleaning with unsupervised learning~\cite{intelTADK}. First it applies deep packet inspection to filter out plain-text flows. Then it groups the remaining encrypted flows by similarity and discards clusters that represent irrelevant traffic. On real world mobile traces, this approach reduces classification accuracy by only 2\% to 2.5\% compared with full manual cleaning.

\section{Overview Design}
\label{sec:overview}

\begin{figure}[!t]
  \centering
  \includegraphics[width=0.5\textwidth]{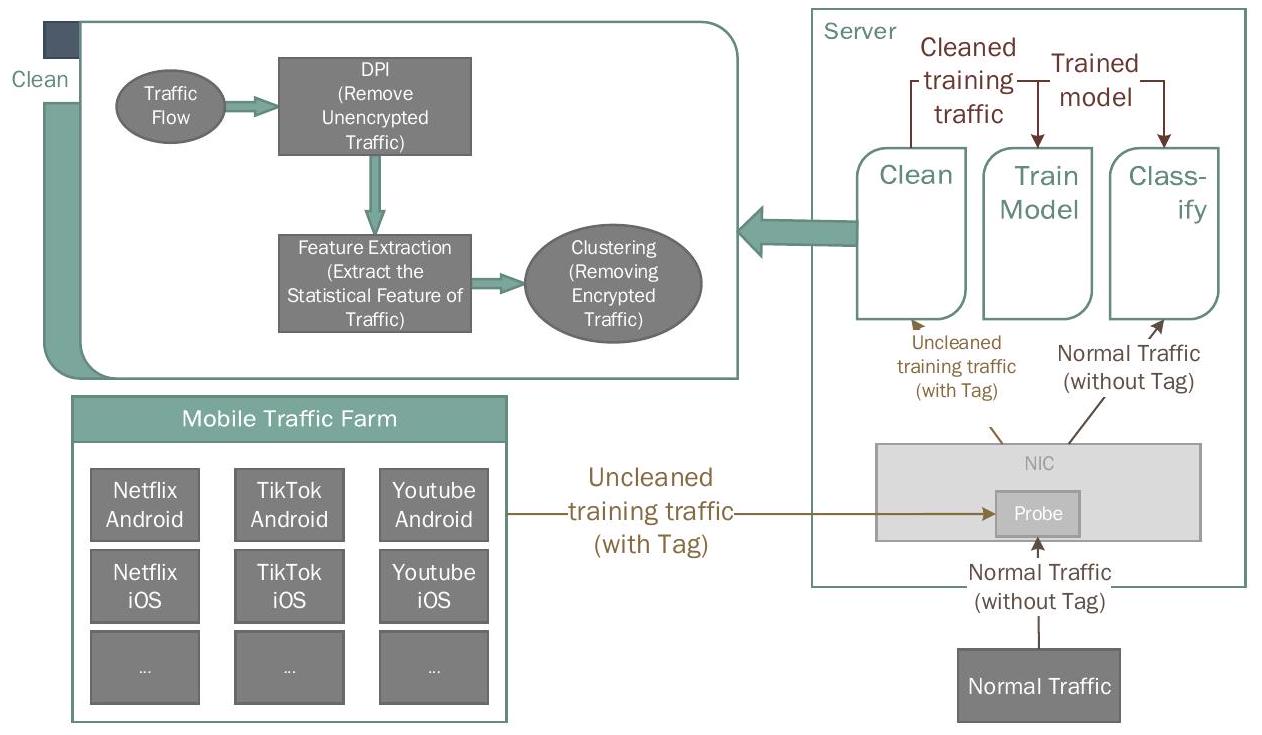}\\
  \caption{The overview of our design. The mobile traffic farm is utilized to generate traffic for every application which is needed for training. We deploy the application in the real smartphone or simulator in the traffic farm.}
  \label{fig:design}
\end{figure}

We build a mobile traffic farm that runs each target app on real devices and simulators (Android and iOS) and tags all flows by MAC address or VLAN ID. This \textit{uncleaned} traffic includes both app‐generated packets and background or control‐plane traffic from the OS. All tagged data is streamed to a server‐side Virtual Network Function: a probe at the NIC feeds the uncleaned traffic into our cleaning module, which filters out irrelevant flows and passes only the cleaned traces to a training service. The resulting model is finally deployed back into the classifier. The overall design as shown in Fig.~\ref{fig:design}.

\begin{figure}[!t]
  \centering
  \includegraphics[width=0.3\textwidth]{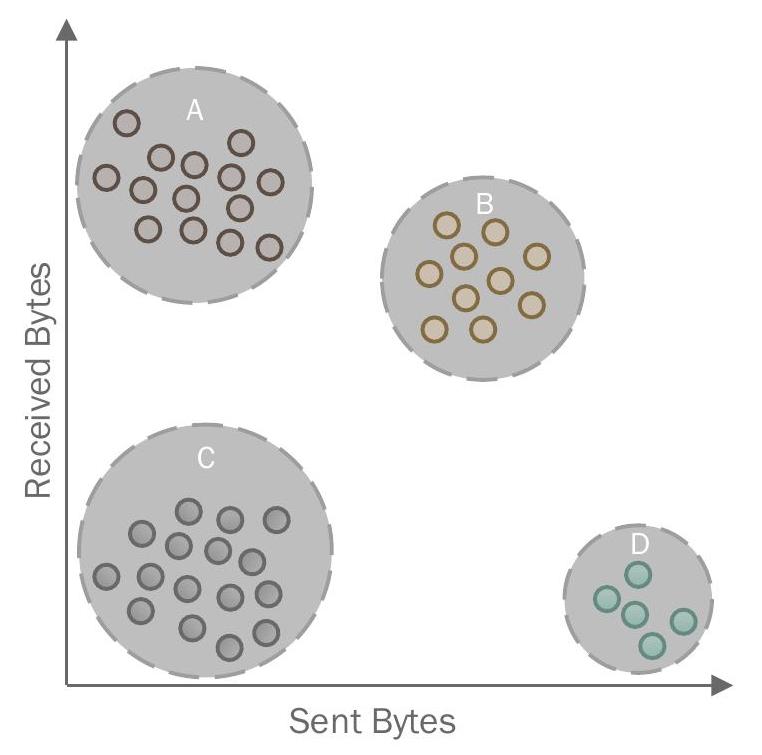}\\
  \caption{An example of utilizing the clustering algorithm in cleaning encrypted traffic. In this figure, we only use two features: Sent Bytes and Received Bytes as an example.}
  \label{fig:cluster}
\end{figure}

\section{Train Model with Cleaned Traffic}
\label{sec:model}
Our cleaning module proceeds in three stages. First, we extract statistical features from each packet flow. Second, we apply DPI~\cite{qiu2021teddy} to discard unencrypted background and service traffic (e.g., DNS, Google/Apple, Cloudflare). Third, we cluster the remaining encrypted flows and remove entire clusters whose feature profiles (e.g., heartbeat traffic or upload‐only flows) do not match our target patterns. The output is a fully cleaned dataset for model training.

In the first two stages, we use feature‐extraction library~\cite{qiu2022traffic} to compute metrics such as BytesIn, BytesOut, PacketsIn, PacketsOut, flow duration and header/payload mean size, producing a structured feature table. Concurrently, our DPI tool filters out unwanted cleartext traffic, yielding a \textit{pre‐cleaned} set of encrypted flows described by their statistical vectors.

Finally, we apply unsupervised clustering, such as K-means for speed or hierarchical clustering for higher accuracy—on six key features: BytesIn, BytesOut, PacketsIn, PacketsOut, Duration and Ratio. Ratio is a value ranging from -1 to 1 and is computed by $\frac{{BytesIn-BytesOut}}{{BytesIn }+{BytesOut}}$. If the ratio is closer to 1, it indicates the traffic is similar to download traffic and vice versa.. Empirical results show that $3 \sim 4$ clusters suffice to isolate noise. For example, selecting the cluster with $ratio > 0.9$ extracts video‐stream data plane traffic, while choosing the cluster with long duration and minimal upload captures heartbeat or control‐plane flows. The retained clusters form the cleaned training. We give an example in Fig.~\ref{fig:cluster}.

\begin{table}
\label{tab:data}
    \centering
    \caption{Dataset}
\begin{center}
\begin{tabular}{|l|l|l|l|l|}
\hline
\rule[-6pt]{0mm}{18pt} Dataset & Size & Number of Flows &  Duration \\
\hline
\rule[-6pt]{0mm}{18pt} Youku & 1.94 GB & 16017 &  2 hours \\
\hline
\rule[-6pt]{0mm}{18pt} Weishi & 1.83 GB & 12097 &  2 hours \\
\hline
\rule[-6pt]{0mm}{18pt} Kuaishou & 1.52 GB & 10148 &  2 hours \\
\hline
\rule[-6pt]{0mm}{18pt} Tiktok & 1.47 GB & 9718 &  2 hours \\
\hline
\rule[-6pt]{0mm}{18pt} Bilibili & 2.02 GB & 14132 &  2 hours \\
\hline
\end{tabular}
\end{center}
\end{table}

\section{Evaluation}
\label{sec:eva}

\subsection{Accuracy}
\begin{figure}[!t]
  \centering
  \includegraphics[width=0.5\textwidth]{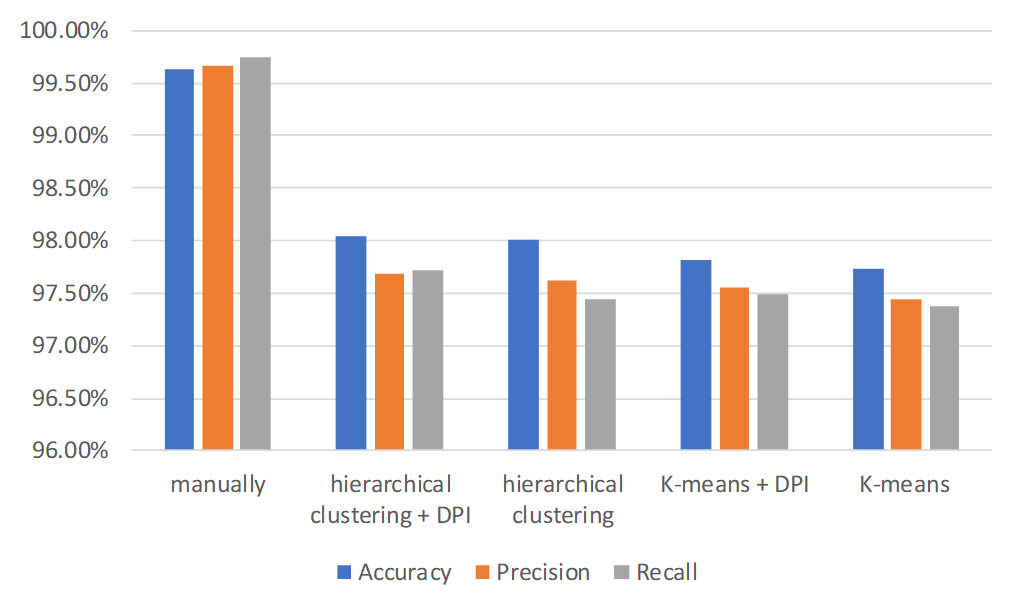}\\
  \caption{The accuracy/precision/recall comparison between manually cleaning and our cleaning framework. We do not show the uncleaned result since the accuracy of uncleaned traffic is less than $50 \%$. We use both K-means and hierarchical clustering algorithms to evaluate traffic cleaning. The result shows that our cleaning framework only has $2 \% \sim 3 \%$ accuracy loss than the manually traffic cleaning.}
  \label{fig:accu}
\end{figure}

We give the detailed information of our dataset in TABLE~I. The traffic data is collected by the mobile traffic farm. 
In order to obtain the accuracy/precision/recall of our cleaning framework, we choose the 75\% traffic for training and 25\% traffic for testing. After performing the clustering algorithm, we use the data plane traffic to train a classifier by choosing the cluster whose traffic ratio is closer to 1. We also manually clean the traffic by checking every packet offline. We use all 5 application traffic to train a multi-class random forest classifier based on scikit-learn. From Fig.~\ref{fig:accu} we can see that utilizing the clustering algorithm only brings $2 \% \sim 2.5 \%$ accuracy loss then the manual traffic cleaning approach. Moreover, the hierarchical clustering algorithm has slightly higher accuracy than the K-means algorithm.

\begin{figure}[!t]
  \centering
  \includegraphics[width=0.5\textwidth]{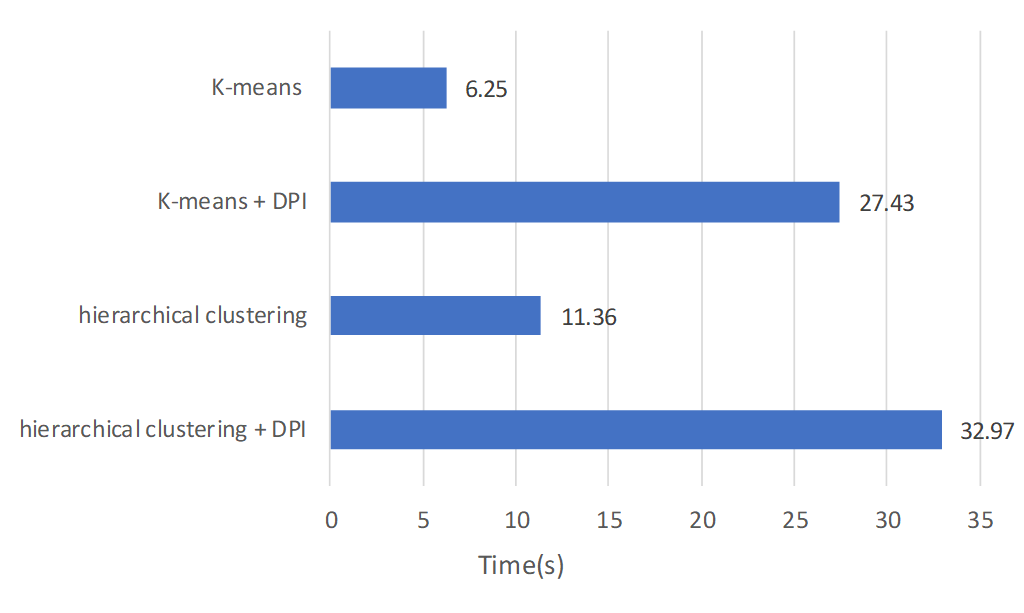}\\
  \caption{The performance comparison between different clustering algorithm. We do not show the time of manually traffic cleaning since it costs more than several hours. The result shows that the hierarchical clustering algorithm uses more time than the K-means algorithm}
  \label{fig:perf}
\end{figure}

\subsection{Performance}
As we have claimed that manually checking every packet is expensive, which may consume over hours or days, we evaluate the time consuming of our traffic cleaning framework to show the computing performance. Fig.~\ref{fig:perf} shows that the time consuming of our framework is less than $33 s$ with DPI, $12 s$ without DPI, which is significantly less than manually traffic cleaning. Moreover, the hierarchical cluster algorithm uses more time than the K-means algorithm.

\bibliographystyle{IEEEtran}
\bibliography{main}
\end{document}